\documentstyle[12pt]{article}

\def\new{\newcommand}
\def\renew{\renewcommand}
\new{\nrao}{naive realism about operators}
\new{\Sc}{Schr\"{o}dinger}
\new{\se}{Schr\"odinger's equation}
\new{\BM}{Bohmian mechanics}
\new{\qf}{quantum formalism}
\new{\qm}{quantum mechanics}
\new{\qt}{quantum theory}
\new{\wf}{wave function}
\new{\ewf}{effective wave function}
\new{\cwf}{conditional wave function}
\new{\qe}{quantum equilibrium}
\new{\C}{{\sf I\!\!\!C}}
\renew{\Delta}{{{\bold \nabla}}^{\;2}}
\new{\be}{\begin{equation}}
\new{\ee}{\end{equation}}
\new{\eq}[1]{(\ref{#1})}
\new{\bold}[1]{\mbox{\boldmath$#1$\unboldmath}}
\renew{\a}{\alpha}
\new{\ot}{\otimes}
\new{\psia}{\psi_{\a}}
\renew{\H}{\mbox{${\cal H}$}}
\renew{\P}{\mbox{${\rm I\!P}$}}
\new{\R}{\mbox{${\rm I\!R}$}}
\new{\E}{\mbox{${\cal E}$}}
\new{\psisq}{|\psi|^2}
\new{\born}{\rho=|\psi|^2}
\begin{document}
\title{Bohmian Mechanics as the Foundation of\\ Quantum Mechanics}
\author{ D. D\"{u}rr\\
Mathematisches Institut der Universit\"{a}t M\"{u}nchen\\
Theresienstra{\ss}e 39, 80333 M\"{u}nchen, Germany
\and
S. Goldstein\\
Department of Mathematics, Rutgers University\\
New Brunswick, NJ 08903, USA
\and
N. Zangh\`{\i}\\
Istituto di Fisica dell'Universit\`a di Genova, INFN \\
via Dodecaneso 33, 16146 Genova, Italy}
\date{November 13, 1995}
\maketitle
\openup2.434\jot

In order to arrive at Bohmian mechanics from standard nonrelativistic
quantum mechanics one need do almost nothing!  One need only complete the
usual quantum description in what is really the most obvious way: by simply
including the positions of the particles of a quantum system as part of the
state description of that system, allowing these positions to evolve in the
most natural way.  The entire quantum formalism, including the uncertainty
principle and quantum randomness, emerges from an analysis of this
evolution. This can be expressed succinctly---though in fact not succinctly
enough---by declaring that the essential innovation of Bohmian mechanics is
the insight that {\it particles move\/}!

\section{Bohmian Mechanics is Minimal}
\setcounter{equation}{0}

%\small\normalsize
\begin{quotation}
\noindent Is it not clear from the smallness of the scintillation on
the screen that we have to do with a particle? And is it not clear, from
the diffraction and interference patterns, that the motion of the particle
is directed by a wave? De Broglie showed in detail how the motion of a
particle, passing through just one of two holes in screen, could be
influenced by waves propagating through both holes.    And so influenced
that the particle does not go where the waves cancel out, but is attracted
to where they cooperate. This idea seems to me so natural and simple, to
resolve the wave-particle dilemma  in such a clear and ordinary way, that
it is a great mystery to me that it was so generally ignored. (Bell 1987, 191)
\end{quotation}

According to orthodox \qt, the {\it complete\/} description of a system of
particles is provided by its \wf.  This statement is somewhat
problematical: If ``particles'' is intended with its usual
meaning---point-like entities whose most important feature is their
position in space---the statement is clearly false, since the complete
description would then have to include these positions; otherwise, the
statement is, to be charitable, vague. \BM\ is the theory that emerges
when we indeed insist that ``particles'' means particles.

According to \BM, the complete description or state of an $N$-particle
system is provided by its \wf\ $\psi(q,t)$, where $q=({\bf q}_1,
\dots ,  {\bf q}_N) \in \R ^{3N},$  {\it and\/} its configuration $ Q=({\bf
Q}_1,
\dots ,  {\bf Q}_N) \in \R ^{3N},$  where the ${\bf Q}_k$  are the positions of
the particles. The \wf, which evolves according to \se,

\be\label{eqsc}i\hbar\frac{\partial \psi }{\partial t}  =
H\psi \,.
\ee
choreographs the motion of the particles: these evolve according to the
equation

\be\label{velo}
\frac{d{\bf Q}_k}{dt} = \frac{\hbar}{m_{k}}\frac{ {\rm
Im}(\psi^*\bold{\nabla}_{\!k}\psi)}{\psi^*\psi}\,({\bf Q}_1,
\dots ,  {\bf Q}_N)
\ee
where $\bold{\nabla}_{\!k}=\partial/\partial \bold q_{\!k}$. In eq.
(\ref{eqsc}),
 $H$ is the usual nonrelativistic \Sc\ Hamiltonian; for
spinless particles it is of the form
\be\label{sh}
H=-{\sum}_{k=1}^{N}
\frac{{\hbar}^{2}}{2m_{k}}\Delta_{\!k}  + V,
\ee
containing as parameters the masses $m_1\dots, m_N$ of the particles as
well as the potential energy function $V$ of the system. For an
$N$-particle system of nonrelativistic particles, equations
(\ref{eqsc}) and (\ref{velo}) form a complete specification of the
theory.$^1$
There is no need, and indeed no room, for any further {\it axioms\/},
describing either the behavior of other observables or the effects of
measurement.

In view of what has so often been said---by most of the leading physicists
of this century and in the strongest possible terms---about the radical
implications of quantum theory, it is not easy to accept that Bohmian
mechanics really works.  However, in fact, it does: Bohmian mechanics
accounts for all of the phenomena governed by nonrelativistic quantum
mechanics, from spectral lines and quantum interference experiments to
scattering theory and superconductivity.  In particular, the usual
measurement postulates of quantum theory, including collapse of the wave
function and probabilities given by the absolute square of probability
amplitudes, emerge as a consequence merely of the two equations of motion
for Bohmian mechanics---Schr\"odinger's equation and the guiding
equation---without the traditional invocation of a special and somewhat
obscure status for observation.

It is important to bear in mind that regardless of which observable we
choose to measure, the result of the measurement can be assumed to be given
configurationally, say by some pointer orientation or by a pattern of ink
marks on a piece of paper. Then the fact that \BM\ makes the same
predictions as does orthodox quantum theory for the results of any
experiment---for example, a measurement of momentum or of a spin
component---at least assuming a random distribution for the configuration
of the system and apparatus at the beginning of the experiment given by
$|\psi(q)|^2$, is a more or less immediate consequence of
(\ref{velo}). This is because the quantum continuity equation
\be\label{quflu}
\frac{\partial |\psi(q,t)|^2}{\partial t} + \hbox{\rm div}\, J^{\psi}(q,t) =0,
\ee
where
\be\label{flux}
J^{\psi}(q,t) = ({\bf J}_1^{\psi}(q,t), \dots, {\bf J}_N^{\psi}(q,t))
\ee
with
\be {\bf J}_k^\psi= \frac{\hbar}{m_k}{\rm Im}\,(\psi^*\bold{\nabla}_{\!k}\psi)
\ee
is the {\it quantum probability current\/}, an equation that is a simple
consequence of \se, becomes the classical continuity
equation
\be\label{clflu}
\frac{\partial \rho}{\partial t} + \hbox{\rm div}\, \rho v =0
\ee
for the system $dQ/dt=v$ defined by (\ref{velo})---the equation governing
 the evolution of a probability density $\rho$ under the motion
defined by the guiding equation (\ref{velo})---when $\rho=|\psi|^2=\psi^*\psi$,
the
{\it quantum equilibrium\/} distribution. In other words, if the
probability density for the configuration satisfies
$\rho(q,t_0)=|\psi(q,t_0)|^2$ at some time $t_0$, then the density to which
this is carried by the motion (\ref{velo}) at any time $t$ is also given by
$\rho(q,t)=|\psi(q,t)|^2$. This is an extremely important property of \BM,
one that expresses a certain compatibility between the two equations of
motion defining the dynamics, a property which we call the {\it
equivariance\/} of the probability distribution $|\psi|^2$. (It of course
holds for any Bohmian system and not just the system-apparatus composite
upon which we have been focusing.)

While the meaning and justification of the {\it quantum equilibrium
hypothesis\/} that $\rho=|\psi|^2$ is a delicate matter, to which we shall
later return, it is important to recognize at this point that, merely as a
consequence of (\ref{velo}), \BM\ is a counterexample to all of the claims
to the effect that a deterministic theory cannot account for quantum
randomness in the familiar statistical mechanical way, as arising from
averaging over ignorance: \BM\ is clearly a deterministic theory, and, as
we have just explained, it does account for quantum randomness as arising
from averaging over ignorance given by $|\psi(q)|^2$.

\bigskip

Note that Bohmian mechanics incorporates Schr\"odinger's equation into a
rational theory, describing the motion of particles, merely by adding a
single equation, the guiding equation (\ref{velo}), a first-order evolution
equation for the configuration. In so doing it provides a precise role for
the \wf\ in sharp contrast with its rather obscure status in orthodox quantum
theory.  Moreover, if we take Schr\"odinger's equation directly into
account---as of course we should if we seek its rational completion---this
additional equation emerges in an almost inevitable manner, indeed via
several routes.  Bell's preference is to observe that the probability current
$J^{\psi}$
and the probability density $\rho=\psi^*\psi$ would
classically be related (as they would for any dynamics given by a first-order
ordinary differential equation) by $J=\rho v$, obviously suggesting that
\be\label{vjr}
dQ/dt=v=J/\rho,
\ee
which is the guiding equation (\ref{velo}).

Bell's route to (\ref{velo}) makes it clear that it does not require great
imagination to arrive at the guiding equation. However, it does not show
that this equation is in any sense mathematically inevitable.  Our own
preference is to proceed in a somewhat different manner, avoiding any use,
even in the motivation for the theory, of probabilistic notions, which are
after all somewhat subtle, and see what symmetry considerations alone might
suggest. Assume for simplicity that we are dealing with spinless
particles. Then one finds (D\"urr, Goldstein and Zangh\`{i} 1992, first
reference) that, given \se, the simplest choice, compatible with overall
Galilean and time-reversal invariance, for an evolution equation for the
configuration, the simplest way a suitable velocity vector can be extracted
from the scalar field $\psi$, is given by \be\label{sd} \frac{d{\bf
Q}_k}{dt}=\frac {\hbar}{m_k}\mbox {Im}\frac{\bold{\nabla}_{\!k}\psi}{\psi},
\ee which is of course equivalent to (\ref{velo}): The $\bold{\nabla}$ on
the right-hand side is suggested by rotation invariance, the $\psi$ in the
denominator by homogeneity---i.e., by the fact that the wave function
should be understood projectively, an understanding required for the
Galilean invariance of \Sc's equation alone---and the ``Im'' by time-reversal
invariance, since time-reversal is implemented on $\psi$ by complex
conjugation, again as demanded by \Sc's equation. The constant in front
is precisely what is required for covariance under Galilean boosts.

\section{Bohmian Mechanics and Classical Physics}

You will no doubt have noticed that the quantum potential, introduced and
emphasized by Bohm (Bohm 1952 and Bohm and Hiley 1993)---but repeatedly
dismissed, by omission, by Bell (Bell 1987)---did not appear in our
formulation of \BM. Bohm, in his seminal (and almost universally ignored!)
1952 hidden-variables paper (Bohm 1952),  wrote the wave function $\psi$
in the polar form $\psi=Re^{iS/\hbar}$ where $S$ is real and $R\ge 0$, and
then rewrote Schr\"odinger's equation in terms of these new variables,
obtaining a pair of coupled evolution equations: the continuity equation
(\ref{clflu}) for $\rho=R^2$, which suggests that $\rho$ be interpreted as
a probability density, and a modified Hamilton-Jacobi equation for $S$,
\be\label{hj}
\frac{\partial S}{\partial t}+H(\nabla S,q)+U=0,
\ee
where $H=H(p,q)$ is the classical Hamiltonian function corresponding to
(\ref{sh}), and
\be\label{qp}
U=-\sum_k \frac{\hbar^2}{2m_k}\frac{{\bold\nabla_{\!k}}^2R}{R}.
\ee
Noting that this equation differs from the usual classical Hamilton-Jacobi
equation only by the appearance of an extra term, the {\it quantum
potential\/} $U$, Bohm then used the equation to define particle
trajectories just as is done for the classical Hamilton-Jacobi equation,
that is, by identifying $\nabla S$ with $mv$, i.e., by
\be\label{bvel}
\frac{d{\bf Q}_k}{dt}=\frac {\bold{\nabla}_{\!k}S}{m_k},
\ee
which is equivalent to (\ref{sd}). The resulting motion is
precisely what would have been obtained classically if the particles were
acted upon by the force generated by the quantum potential in addition to
the usual forces.

Bohm's rewriting of Schr\"odinger's equation via variables that seem
interpretable in classical terms does not come without a cost. The most
obvious cost is increased complexity: Schr\"odinger's equation is rather
simple, not to mention linear, whereas the modified Hamilton-Jacobi
equation is somewhat complicated, and highly nonlinear---and still requires
the continuity equation for its closure. The quantum potential itself is
neither simple nor natural [even to Bohm it has seemed ``rather strange and
arbitrary'' (Bohm 1980, 80)] and it is not very satisfying to think of the
quantum revolution as amounting to the insight that nature is classical
after all, except that there is in nature what appears to be a rather ad
hoc additional force term, the one arising from the quantum potential.

Moreover, the connection between classical mechanics and Bohmian mechanics
that is suggested by the quantum potential is rather misleading.  Bohmian
mechanics is not simply classical mechanics with an additional force term. In
Bohmian mechanics the velocities are not independent of positions, as they
are classically, but are constrained by the guiding equation
\be\label{ge}
{\bf v}_k=\bold{\nabla}_{\!k} S/m_k.
\ee
In classical Hamilton-Jacobi
theory we also have this equation for the velocity, but there the
Hamilton-Jacobi function $S$ can be entirely eliminated and the description
in terms of $S$ simplified and reduced to a finite-dimensional description,
with basic variables the positions and momenta of all the particles, given
by Hamilton's or Newton's equations.

We wish to stress that since the dynamics for Bohmian mechanics
is  completely defined by Schr\"odinger's equation together with the
guiding equation, there is neither need nor room for any further {\it
axioms\/} involving the quantum potential! Thus the quantum potential
should not be regarded as fundamental, and we should not allow it to
obscure, as it all too easily tends to do, the most basic structure
defining Bohmian mechanics.

We believe that the most serious flaw in the quantum potential formulation
of Bohmian mechanics is that it gives a completely wrong impression of the
lengths to which we must go in order to convert orthodox quantum theory
into something more rational.$^2$ The quantum potential suggests, and
indeed it has often been stated, that in order to transform Schr\"odinger's
equation into a theory that can account, in what are often called
``realistic'' terms, for quantum phenomena, many of which are dramatically
nonlocal, we must incorporate into the theory a quantum potential of a
grossly nonlocal character.

We have already indicated why such sentiments are inadequate, but we would
like to go further. \BM\ should be regarded as a first-order theory, in
which it is the velocity, the rate of change of position, that is
fundamental in that it is this quantity that is specified by the theory,
directly and simply, with the second-order (Newtonian) concepts of
acceleration and force, work and energy playing no fundamental role. {}From
our perspective the artificiality suggested by the quantum potential is the
price one pays if one insists on casting a highly nonclassical theory into
a classical mold.

This is not to say that these second-order concepts play {\it no\/} role in
\BM; they are emergent notions, fundamental to the theory to which \BM\
converges in the ``classical limit,'' namely, Newtonian mechanics.
Moreover, in order most simply to see that Newtonian mechanics should be
expected to emerge in this limit, it is convenient to transform the
defining equations (\ref{eqsc}) and (\ref{velo}) of \BM\ into Bohm's
Hamilton-Jacobi form. One then sees that the (size of the) quantum
potential provides a rough measure of the deviation of \BM\ from its
classical approximation.

It might be objected that mass is also a second-order concept, one that most
definitely does play an important role in the very formulation of \BM. In
this regard we would like to make several comments. First of all, the
masses appear in the basic equations only in the combination
$m_k/\hbar\equiv\mu_k$. Thus eq. (\ref{velo}) could more efficiently be
written as
\be\label{evelo}
\frac{d{\bf Q}_k}{dt} = \frac{1}{\mu_{k}}\frac{ {\rm
Im}(\psi^*\bold{\nabla}_{\!k}\psi)}{\psi^*\psi}\,,
\ee
and if we divide \se\ by $\hbar$ it assumes the form
\be\label{ese}
i\frac{\partial \psi }{\partial t}=-{\sum}_{k=1}^{N}
\frac{1}{2\mu_{k}}\Delta_{\!k}\psi  + \hat V\psi,
\ee
with $\hat V=V/\hbar$. Thus it seems more appropriate to regard the
naturalized masses $\mu_k$, which in fact have the dimension of
[time]/[length]$^2$, rather than the original masses $m_k$, as the
fundamental parameters of the theory. Notice that if naturalized parameters
(including also naturalized versions of the other coupling constants such
as the naturalized electric charge $\hat e=e/\sqrt\hbar$) are used,
Planck's constant $\hbar$ disappears from the formulation of this quantum
theory. Where $\hbar$ remains is merely in the equations $m_k=\hbar\mu_k$
and $e^2=\hbar {\hat e}^2$ relating the parameters---the masses and the
charges---in the natural microscopic units with those in the natural units
for the macroscopic scale, or, more precisely, for the theory, Newtonian
mechanics, that emerges on this scale.

It might also be objected that notions such as {\it inertial\/} mass and
the quantum potential are necessary if \BM\ is to provide us with any sort
of {\it intuitive \/} explanation of quantum phenomena, i.e., explanation
in familiar terms, presumably such as those involving only the concepts of
classical mechanics.  (See, for example, the contribution of Baublitz and
Shimony to this volume.) It hardly seems necessary to remark, however, that
physical explanation, even in a realistic framework, need not be in terms
of classical physics.

Moreover, when classical physics was first propounded by Newton, this
theory, invoking as it did action-at-a-distance, did not provide an
explanation in familiar terms. Even less intuitive was Maxwell's
electrodynamics, insofar as it depended upon the reality of the
electromagnetic field. We should recall in this regard the lengths to which
physicists, including Maxwell, were willing to go in trying to provide an
intuitive explanation for this field as some sort of disturbance in a
material substratum to be provided by the Ether. These attempts of course
failed, but even had they not, the success would presumably have been
accompanied by a rather drastic loss of mathematical simplicity.  In the
present century fundamental physics has moved sharply away from the search
for such intuitive explanations in favor of explanations having an air of
mathematical simplicity and naturalness, if not inevitability, and this has
led to an astonishing amount of progress. It is particularly important to
bear these remarks on intuitive explanation in mind when we come to the
discussion in the next section of the status of quantum observables,
especially spin.

The problem with orthodox quantum theory is not
that it is unintuitive. Rather the problem is that

\begin{quotation} \noindent ...conventional formulations of quantum theory,
and of quantum field theory in particular, are unprofessionally vague and
ambiguous. Professional theoretical physicists ought to be able to do
better. Bohm has shown us a way. (Bell 1987, 173)
\end{quotation}
The problem, in other words, with orthodox quantum theory is not that it
fails to be intuitively formulated, but rather that, with its incoherent
babble about measurement, it is not even well formulated!

\section{What about Quantum Observables?}

We have argued that quantities such as mass do not have the same meaning in
\BM\ as they do classically. This is not terribly surprising if we bear in
mind that the meaning of theoretical entities is ultimately determined by
their role in a theory, and thus when there is a drastic change of theory,
a change in meaning is almost inevitable. We would now like to argue that
with most observables, for example energy and momentum, something much more
dramatic occurs: In the transition from classical mechanics they cease to
remain properties at all. Observables, such as spin, that have no classical
counterpart also should not be regarded as properties of the system. The
best way to understand the status of these observables---and to better
appreciate the minimality of Bohmian mechanics---is Bohr's way: What are
called quantum observables obtain meaning {\em only} through their
association with specific {\em experiments}. We believe that Bohr's point
has not been taken to heart by most physicists, even those who regard
themselves as advocates of the Copenhagen interpretation, and that the
failure to appreciate this point nourishes a kind of naive realism about
operators, an uncritical identification of operators with properties, that
is the source of most, if not all, of the continuing confusion concerning
the foundations of quantum mechanics.

Information about a system does not spontaneously pop into our heads, or
into our (other) ``measuring'' instruments; rather, it is generated by an
{\it experiment:\/} some physical interaction between the system of
interest and these instruments, which together (if there is more than one)
comprise the {\it apparatus\/} for the experiment.  Moreover, this
interaction is defined by, and must be analyzed in terms of, the physical
theory governing the behavior of the composite formed by system and
apparatus. If the apparatus is well designed, the experiment should somehow
convey significant information about the system. However, we cannot hope to
understand the significance of this ``information''---for example, the
nature of what it is, if anything, that has been measured---without some
such theoretical analysis.

Whatever its significance, the information conveyed by the experiment is
registered in the apparatus as an {\it output\/}, represented, say, by the
orientation of a pointer. Moreover, when we speak of an experiment, we have
in mind a fairly definite initial state of the apparatus, the ready state,
one for which the apparatus should function as intended, and in particular
one in which the pointer has some ``null'' orientation, say as in Figure 1.
\setlength{\unitlength}{.2pt}
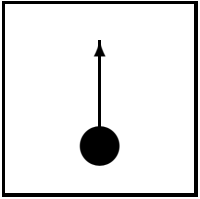
\begin{figure}
\begin{picture}(360,360)(-920,0)
\thicklines
\put(0,0){\framebox(360,360)}
\put(180,90){\vector(0,1){200}}
\put(180,90){\circle*{100}}
\end{picture}
\caption{Initial setting of apparatus.}
\end{figure}

For \BM\ we should expect in general that, as a consequence of the \qe\
hypothesis, the justification of which we shall address in Section 4 and
which we shall now simply take as an assumption, the
outcome of the experiment---the final pointer orientation---will be random:
Even if the system and apparatus initially have definite, known \wf s, so
that the outcome is determined by the initial configuration of system and
apparatus, this configuration is random, since the composite system is in
\qe, i.e., the distribution of this configuration  is given by $|\Psi(x,y)|^2$,
where $\Psi$ is the \wf\ of the system-apparatus composite and $x$ respectively
$y$
is the generic system respectively apparatus configuration. There are, however,
special experiments whose outcomes are somewhat less random than we might
have thought possible.

In fact, consider a {\it measurement-like\/} experiment, one which is {\it
reproducible\/} in the sense that it will yield the same outcome as
originally obtained if it is immediately repeated. (Note that this means
that the apparatus must be immediately reset to its ready state, or a
fresh apparatus must be employed, while the system is not tampered with so
that its initial state for the repeated experiment is its final state
produced by the first experiment.) Suppose that this experiment admits,
i.e., that the apparatus is so designed that there are, only a finite (or
countable) number of possible outcomes $\a$,$^3$ for example,
$\a=$``left'' and $\a=$``right'' as in Figure 2.
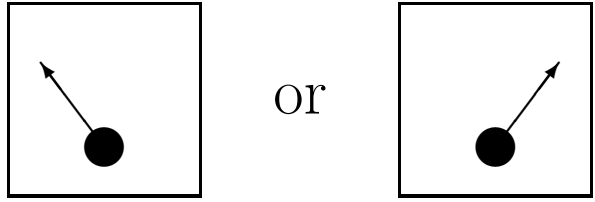
\begin{figure}
\begin{picture}(1100,360)(-550,0)
\put(0,0)
{%
\begin{picture}(360,360)
\thicklines
\put(0,0){\framebox(360,360)}
\put(180,90){\vector(-3,4){120}}
\put(180,90){\circle*{100}}
\end{picture}%
}
\put(360,0){\makebox(380,360){\Huge or}}
\put(740,0)
{%
\begin{picture}(360,360)
\thicklines
\put(0,0){\framebox(360,360)}
\put(180,90){\vector(3,4){120}}
\put(180,90){\circle*{100}}
\end{picture}%
}
\end{picture}
\caption{Final apparatus readings.}
\end{figure}
\noindent
The experiment also usually comes equipped with a {\it calibration\/}
$\lambda_\a$, an assignment of numerical values (or a vector of such
values) to the various outcomes $\a$.

It can be shown (Daumer, D\"urr, Goldstein and Zangh\`{i} 1996), under
further simplifying assumptions, that for such reproducible experiments
there are special subspaces $\H_\a$ of the system Hilbert space $\H$ of
(initial) \wf s, which are mutually orthogonal and span the entire system
Hilbert space
\be\label{Hdecomp}
\H=\bigoplus_\a{\H_\a},
\ee
such that if the system's \wf\ is initially in $\H_\a$, outcome $\a$
definitely occurs and the value $\lambda_\a$ is thus definitely obtained.
It then follows that for a general initial system \wf
\be\label{psidecomp}
\psi=\sum_\a\psi_\a\equiv\sum_\a{P_{\H_\a}}\psi
\ee
where $P_{\H_\a}$ is the projection onto the subspace $\H_\a$, the outcome
$\a$ is obtained with (the usual) probability$^4$
\be\label{prob}
p_\a={\Vert P_{\H_\a}\psi\Vert}^2.
\ee
In particular, the expected value obtained is
\be\label{ev}
\sum_\a{p_\a \lambda_\a}=\sum_\a{\lambda_\a{\Vert
 P_{\H_\a}\psi\Vert}^2}=\langle\psi,A\psi\rangle
\ee
where
\be\label{A}
A=\sum_{\a}{\lambda_\a P_{\H_\a}}
\ee
and $\langle\,\cdot\,,\,\cdot\,\rangle$ is the usual inner product:
\be\label{ip}
\langle\psi,\phi\rangle=\int{{\psi}^*(x)\,\phi(x)\,dx}.
\ee

What we wish to emphasize here is that, insofar as the statistics for the
values which result from the experiment are concerned, the relevant data
for the experiment are the collection ($\H_\a$) of special subspaces,
together with the corresponding calibration ($\lambda_\a$), and {\it this data
is compactly expressed and represented by the self-adjoint operator $A$, on
the system Hilbert space $\H$, given by \eq{A}.\/} Thus with a reproducible
experiment $\E$ we naturally associate an operator $A=A_{\E}$,
\be\label{EA}
\E\mapsto A,
\ee
a single mathematical object, defined on the system alone, in terms of
which an efficient description of the possible results is achieved. If we
wish we may speak of operators as observables, but if we do so it is
important that we appreciate that in so speaking we merely refer to what we
have just sketched: the role of operators in the description of certain
experiments.$^5$

In particular, so understood the notion of operator-as-observable in no way
implies that anything is measured in the experiment, and certainly not the
operator itself!  In a general experiment no system property is being
measured, even if the experiment happens to be measurement-like. Position
measurements are of course an important exception. What in general is going
on in obtaining outcome $\alpha$ is completely straightforward and in no
way suggests, or assigns any substantive meaning to, statements to the
effect that, prior to the experiment, observable $A$ somehow had a value
$\lambda_\a$---whether this be in some determinate sense or in the sense of
Heisenberg's ``potentiality'' or some other ill-defined fuzzy sense---which
is revealed, or crystallized, by the experiment.$^6$

Much of the preceding sketch of the emergence and role of operators as
observables in \BM, including of course the von Neumann-type picture of
``measurement'' at which we arrive, applies as well to orthodox
\qt.$^7$ In fact, it would appear that the argument against \nrao\ provided
by such an analysis has even greater force from an orthodox perspective:
Given the initial \wf, at least in \BM\ the outcome of the particular
experiment is determined by the initial configuration of system and
apparatus, while for orthodox \qt\ there is nothing in the initial state
which completely determines the outcome. Indeed, we find it rather
surprising that most proponents of the von Neumann analysis of measurement,
beginning with von Neumann, nonetheless seem to retain their \nrao. Of
course, this is presumably because more urgent matters---the measurement
problem and the suggestion of inconsistency and incoherence that it
entails---soon force themselves upon one's attention. Moreover such
difficulties perhaps make it difficult to maintain much confidence about
just what {\it should\/} be concluded from the ``measurement'' analysis,
while in \BM, for which no such difficulties arise, what should be
concluded is rather obvious.

\bigskip

It might be objected that we are claiming to arrive at the \qf\ under
somewhat unrealistic assumptions, such as, for example, reproducibility.
(We note in this regard that many more experiments than those satisfying our
assumptions can be associated with operators in exactly the manner we have
described.) We agree. But this objection misses the point of the exercise.
The \qf\ itself is an idealization; when applicable at all, it is only as
an approximation. Beyond illuminating the role of operators as ingredients
in this formalism, our point was to indicate how naturally it emerges.  In
this regard we must emphasize that the following question arises for
quantum orthodoxy, but does not arise for \BM: For precisely which theory is
the \qf\ an idealization?

That the \qf\ is merely an idealization, rarely directly relevant in
practice, is quite clear. For example, in the real world the projection
postulate---that when the measurement of an observable yields a specific
value, the \wf\ of the system is replaced by its projection onto the
corresponding eigenspace---is rarely satisfied. More important, a great
many significant real-world experiments are simply not at all associated
with operators in the usual way. Consider for example an electron with
fairly general initial \wf, and surround the electron with a
``photographic'' plate, away from (the support of the \wf\ of) the
electron, but not too far away. This set-up measures the position of
``escape'' of the electron from the region surrounded by the plate. Notice
that since in general the time of escape is random, it is not at all
clear which operator should correspond to the escape position---it should
not be the Heisenberg position operator at a specific time, and a
Heisenberg position operator at a random time has no meaning. In fact,
there is presumably no such operator, so that for the experiment just
described the probabilities for the possible results cannot be expressed in
the form \eq{prob}, and in fact are not given by the spectral measure for any
operator.

Time measurements, for example escape times or decay times, are
particularly embarrassing for the \qf. This subject remains mired in
controversy, with various research groups proposing their own favorite
candidates for the ``time operator'' while paying little attention to the
proposals of the other groups. For an analysis of time measurements within
the framework of \BM, see Daumer, D\"urr, Goldstein and Zangh\`{i} 1994 and
the contribution of Leavens to this volume.

Because of such difficulties, it has been proposed (Davies 1976) that we
should go beyond operators-as-observables, to ``generalized observables,''
described by mathematical objects even more abstract than operators. The
basis of this generalization lies in the observation that, by the spectral
theorem, the concept of self-adjoint operator is completely equivalent to
that of (a normalized) projection-valued measure (PV), an
orthogonal-projection-valued additive set function, on the value space
$\R$. Since orthogonal projections are among the simplest examples of
positive operators, a natural generalization of a ``quantum observable'' is
provided by a (normalized) positive-operator-valued measure (POV). (When a
POV is sandwiched by a \wf, as on the right-hand side of \eq{ev}, it
generates a probability distribution.)

It may seem that we would regard this development as a step in the wrong
direction, since it supplies us with a new, much larger class of abstract
mathematical entities about which to be naive realists. But for \BM\ POV's
form an extremely natural class of objects to associate with experiments.
In fact, consider a general experiment---beginning, say, at time 0 and ending
at time $t$---with no assumptions about reproducibility or anything else.
The experiment will define the following sequence of maps: $$\psi \mapsto
\Psi = \psi \otimes \Phi_0 \mapsto \Psi_t
\mapsto d\mu =
\Psi_t^{*} \Psi_t dq \mapsto \tilde{\mu} := \mu \circ F^{-1}$$
Here $\psi$ is the initial \wf\ of the system, and $\Phi_0$ is the initial
\wf\ of the apparatus; the latter is of course fixed, defined by the
experiment. The second map corresponds to the time evolution arising from
the interaction of the system and apparatus, which yields the
\wf\ of the composite system after the experiment, with which we
associate its \qe\ distribution $\mu$, the distribution of the
configuration $Q_t$ of the system and apparatus after the experiment. At
the right we arrive at the probability distribution induced by a function
$F$ from the configuration space of the composite system to some value
space, e.g., $\R$, or ${\R}^m$, or what have you: $\tilde{\mu}$ is
the distribution of $F(Q_t)$. Here $F$ could be completely general, but for
application to the results of real-world experiments $F$ might represent
the ``orientation of the apparatus pointer'' or some coarse-graining
thereof.

Notice that the composite map defined by this sequence, from \wf s to
probability distributions on the value space, is ``bilinear'' or
``quadratic,'' since the middle map, to the \qe\ distribution, is obviously
bilinear, while all the other maps are linear, all but the second trivially
so. Now by elementary functional analysis, the notion of such a bilinear
map is completely equivalent to that of a POV! Thus the emergence and role
of POV's as ``generalized observables'' in \BM\ is merely an expression of
the bilinearity of \qe\ together with the linearity of \Sc's evolution.
Thus the fact that with every experiment is associated a POV, which forms a
compact expression of the statistics for the possible results, is a near
mathematical triviality. It is therefore rather dubious that the occurrence
of POV's as observables---the simplest case of which is that of PV's---can
be regarded as suggesting any deep truths about reality or about
epistemology.

\bigskip

The canonical example of a ``quantum measurement'' is provided by the
Stern-Gerlach experiment. We wish to focus on this example here in order to
make our previous considerations more concrete, as well as to present some
further considerations about the ``reality'' of operators-as-observables.
We wish in particular to comment on the status of spin. We shall therefore
consider a Stern-Gerlach ``measurement'' of the spin of an electron, even
though such an experiment is unphysical (Mott 1929), rather than of the
internal angular momentum of a neutral atom.

We must first explain how to incorporate spin into \BM. This is very easy;
we need do, in fact, almost nothing: Our derivation of \BM\ was based in
part on rotation invariance, which requires in particular that rotations
act on the value space of the \wf.  The latter is rather inconspicuous for
spinless particles---with complex-valued \wf s, what we have been
considering up till now---since rotations then act in a trivial manner on
the value space $\C$. The simplest nontrivial (projective)
representation of the rotation group is the $2$-dimensional, ``spin $\frac
1 2$'' representation; this representation leads to a \BM\ involving
spinor-valued \wf s for a single particle and spinor-tensor-product-valued
\wf\ for many particles.  Thus the \wf\ of a single spin $\frac 1 2$
particle has two components
\be\label{sp}
\psi(\bold q)=\left(\begin{array}{c} \psi_1(\bold q)\\
\psi_2(\bold q)\end{array}\right),
\ee
which get mixed under rotations according to the action generated by the
Pauli spin matrices $\bold \sigma=(\sigma_x,\sigma_y,\sigma_z)$, which
may be taken to be
\be\label{pm}
\sigma_x=\left(\begin{array}{cc}
0&1\\1&0\end{array}\right)\quad\sigma_y=\left(\begin{array}{cc}
0&-i\\i&0\end{array}\right)\quad\sigma_z=\left(\begin{array}{cc}
 1&\;0\\0&-1\end{array}\right)
\ee

Beyond the fact that the \wf\ now has a more abstract value space, nothing
changes from our previous description: The \wf\ evolves via (\ref{eqsc}),
where now the Hamiltonian $H$ contains the Pauli term, for a single
particle proportional to $\bold B\cdot\bold \sigma$, which represents the
coupling between the ``spin'' and an external magnetic field $\bold B$. The
configuration evolves according to (\ref{velo}), with the products of
spinors now appearing there understood as spinor-inner-products.

Let's focus now on a Stern-Gerlach ``measurement of $A=\sigma_z$.'' An
inhomogeneous magnetic field is established in a neighborhood of the
origin, by means of a suitable arrangement of magnets. This magnetic field
is oriented more or less in the positive $z$-direction, and is increasing
in this direction. We also assume that the arrangement is invariant under
translations in the $x$-direction, i.e., that the geometry does not depend
upon $x$-coordinate. An electron, with a fairly definite momentum, is
directed towards the origin along the negative $y$-axis. Its passage
through the inhomogeneous field generates a vertical deflection of its wave
function away from the $y$-axis, which for
\BM\ leads to a similar deflection of the electron's trajectory. If its \wf\
$\psi$ were initially an eigenstate of $\sigma_z$ of eigenvalue $1$ ($-1$),
i.e., if it were of the form$^8$
\be\label{updown}
%% FOLLOWING LINE CANNOT BE BROKEN BEFORE 80 CHAR
\psi=|\uparrow\,\rangle\otimes\phi_0\quad(\psi=|\downarrow\,\rangle\otimes\phi_0)
\ee
where
\be\label{sb}
|\uparrow\,\rangle=\left(\begin{array}{c}
  1\\0\end{array}\right)\quad\mbox{and}\quad|\downarrow\,
\rangle=\left(\begin{array}{c} 0\\1\end{array}\right),
\ee
then the deflection would be in the positive (negative) $z$-direction (by a
rather definite angle).  For a more general initial \wf,  passage
through the magnetic field will, by linearity, split the \wf\ into an
upward-deflected piece (proportional to $|\uparrow\,\rangle$) and a
downward-deflected piece (proportional to $|\downarrow\,\rangle$), with
corresponding deflections of the possible trajectories.

The outcome is registered by detectors placed in the way of these two
``beams.'' Thus of the four kinematically possible outcomes (``pointer
positions'') the occurrence of no detection defines the null output,
simultaneous detection is irrelevant ( since it does not occur if the
experiment is performed one particle at a time), and the two relevant
outcomes correspond  to registration by either the upper or the
lower detector. Thus the calibration for a measurement of $\sigma_z$ is
$\lambda_{\mbox{up}}=1$ and $\lambda_{\mbox{down}}=-1$ (while for a
measurement of the $z$-component of the spin angular momentum itself the
calibration is the product of what we have just described by $\frac 12\hbar$).

Note that one can completely understand what's going on in this
Stern-Gerlach experiment without invoking any additional property of the
electron, e.g., its {\it actual\/} $z$-component of spin that is revealed
in the experiment. For a general initial \wf\ there is no such property;
what is more, the transparency of the analysis of this experiment makes it
clear that there is nothing the least bit remarkable (or for that matter
``nonclassical'') about the {\it nonexistence\/} of this property. As we
emphasized earlier, it is \nrao, and the consequent failure to pay
attention to the role of operators as observables, i.e., to precisely what
we should mean when we speak of measuring operator-observables, that
creates an impression of quantum peculiarity.

Bell has said that (for \BM) spin is not real. Perhaps he should better
have said: {\it ``Even\/} spin is not real,'' not merely because of all
observables, it is spin which is generally regarded as quantum mechanically
most paradigmatic, but also because spin is treated in orthodox \qt\ very
much like position, as a ``degree of freedom''---a discrete index which
supplements the continuous degrees of freedom corresponding to
position---in the \wf.  Be that as it may, his basic meaning is, we
believe, this: Unlike position, spin is not {\it primitive\/},$^9$ i.e., no
{\it actual\/} discrete degrees of freedom, analogous to the {\it actual\/}
positions of the particles, are added to the state description in order to
deal with ``particles with spin.''  Roughly speaking, spin is {\it
merely\/} in the \wf.  At the same time, as just said, ``spin
measurements'' are completely clear, and merely reflect the way spinor \wf
s are incorporated into a description of the motion of configurations.

It might be objected that while spin may not be primitive, so that the result
of our ``spin measurement'' will not reflect any initial primitive property
of the system, nonetheless this result {\it is\/} determined by the initial
configuration of the system, i.e., by the position of our electron,
together with its initial \wf, and as such---as a function
$X_{\sigma_z}(\bold q, \psi)$ of the state of
the system---it is  some property of the system and in particular it is
surely real. Concerning this, several comments.

The function $X_{\sigma_z}(\bold q, \psi)$, or better the property it
represents, is (except for rather special choices of $\psi$) an extremely
complicated function of its arguments; it is not ``natural,'' not a
``natural kind'': It is not something in which, in its own right, we should
be at all interested, apart from its relationship to the {\it result\/} of
this particular experiment.

Be that as it may, it is not even possible to identify this function
$X_{\sigma_z}(\bold q, \psi)$ with the measured spin component, since
different experimental setups for ``measuring the spin component'' may lead
to entirely different functions. In other words $X_{\sigma_z}(\bold q,
\psi)$ is an abuse of notation, since the function $X$ should be labeled,
not by $\sigma_z$, but by the particular experiment for ``measuring
$\sigma_z$''.

For example (Albert 1992, 153), if $\psi$ and the magnetic field have
sufficient reflection symmetry with respect to a plane between the poles of
our SG magnet, and if the magnetic field is reversed, then the sign of what
we have called $X_{\sigma_z}(\bold q,\psi)$ will be reversed: for both
orientations of the magnetic field the electron cannot cross the plane of
symmetry and hence if initially above respectively below the symmetry plane
it remains above respectively below it. But when the field is reversed so
must be the calibration, and what we have denoted by $X_{\sigma_z}(\bold
q,\psi)$ changes sign with this change in experiment. (The change in
experiment proposed by Albert is that ``the {\it hardness box\/} is {\it
flipped over\/}.'' However, with regard to spin this change will
produce essentially no change in $X$ at all. To obtain the reversal of sign,
either the polarity or the geometry of the SG magnet must be reversed, but
not both.)

In general $X_A$ does not exist, i.e., $X_{\E}$, the result of the
experiment $\E$, in general depends upon $\E$ and not just upon $A=A_{\E}$,
the operator associated with $\E$. In foundations of \qm\ circles this
situation is referred to as {\it contextuality,\/} but we believe that this
terminology, while quite appropriate, somehow fails to convey with
sufficient force the rather definitive character of what it entails:
Properties which are merely contextual are not properties at all; they
do not exist, and their failure to do so is in the strongest sense
possible! We thus believe that contextuality reflects little more than the
rather obvious observation that the result of an experiment should depend
upon how it is performed!

\section{The Quantum Equilibrium Hypothesis}

The predictions of \BM\ for the results of a quantum experiment involving a
system-apparatus composite having \wf\ $\psi$ are precisely those of the
quantum formalism, and moreover the quantum formalism of operators as
observables emerges naturally and simply from \BM\ as the very expression
of its empirical import, {\it provided\/} it is assumed that prior to the
experiment the configuration of the system-apparatus composite is random,
with distribution given by $\born$.  But how, in this deterministic theory,
does randomness enter? What is special about $\born$? What exactly does
$\born$ mean---to precisely which ensemble does this probability
distribution refer? And why should $\born$ be true?

We have already said that what is special about the quantum equilibrium
distribution $\born$ is that it is equivariant [see below eq.
(\ref{clflu})], a notion extending that of stationarity to the Bohmian
dynamics (\ref{velo}), which is in general explicitly time-dependent. It is
tempting when trying to justify the use of a particular ``stationary''
probability distribution $\mu$ for a dynamical system, such as the quantum
equilibrium distribution for Bohmian mechanics, to argue that this
distribution has a dynamical origin in the sense that even if the initial
distribution $\mu_0$ were different from $\mu$, the dynamics generates a
distribution $\mu_t$ which changes with time in such a way that $\mu_t$
approaches $\mu$ as $t$ approaches $\infty$ (and that $\mu_t$ is
approximately equal to $\mu$ for $t$ of the order of a ``relaxation
time''). Such `convergence to equilibrium' results---associated with the
notions of `mixing' and `chaos'---are mathematically quite interesting.
However, they are also usually very difficult to establish, even for rather
simple and, indeed, artificially simplified dynamical systems. One of the
nicest and earliest results along these lines, though for a rather special
Bohmian model, is due to Bohm (Bohm 1953).$^{10}$

However, the justification of the \qe\ hypothesis is a problem that by its
very nature can be adequately addressed only on the universal level. To
better appreciate this point, one should perhaps reflect upon the fact that
the same thing is true for the related problem of understanding the origin
of thermodynamic nonequilibrium (!) and irreversibility. As Feynman has
said (Feynman, Leighton and Sands 1963, 46--8),

\begin{quote} Another delight of our subject of physics is that even simple
and idealized things, like the ratchet and pawl, work only because they are
part of the universe.  The ratchet and pawl  works in only one direction
because it has some ultimate contact with the rest of the universe. \dots
its one-way behavior is tied to the one-way behavior of the entire universe.
\end{quote}

\noindent An argument establishing the convergence to \qe\ for local
systems, if it is not part of an argument explaining universal quantum
equilibrium, would leave open the possibility that conditions of local
equilibrium would tend to be overwhelmed, on the occasions when they do
briefly obtain, by interactions with an ambient universal nonequilibrium. In
fact, this is precisely what does happen with thermodynamic equilibrium. In
this regard, it is important to bear in mind that while we of course live
in a universe that is not in universal thermodynamic equilibrium, a fact
that is crucial to everything we experience, all available evidence
supports universal quantum equilibrium. Were this not so, we should expect
to be able to achieve violations of the quantum formalism---even for small
systems. Indeed, we might expect the violations of universal quantum
equilibrium to be as conspicuous as those of thermodynamic equilibrium.

Moreover,  there are some crucial subtleties here, which we can begin to
appreciate by first asking the question: Which systems should be governed
by \BM? The systems which we normally consider are subsystems of a larger
system---for example, the universe---whose behavior (the behavior of the
whole) determines the behavior of its subsystems (the behavior of the
parts). Thus for a Bohmian universe, it is only the universe itself which a
priori---i.e., without further analysis---can be said to be governed by
\BM.

So let's consider such a universe. Our first difficulty immediately
emerges: In practice $\born$ is applied to (small) subsystems.  But only
the universe has been assigned a \wf, which we shall denote by
$\Psi$. What is meant then by the right hand side of $\born$, i.e., by the
\wf\ of a subsystem?

Fix an initial \wf\ $\Psi_0$ for this universe. Then
since the Bohmian evolution is completely deterministic, once the initial
configuration $Q$ of this universe is also specified, all future events,
including of course the results of measurements, are determined. Now let
$X$ be some subsystem variable---say the configuration of the subsystem at
some time $t$---which we would like to be governed by $\born$.  How can
this possibly be, when there is nothing at all random about $X$?

 Of course, if we allow the initial universal configuration $Q$ to be
random, distributed according to the \qe\ distribution ${|\Psi_0(Q)|}^2$,
it follows from equivariance that the universal configuration $Q_t$ at
later times will also be random, with distribution given by ${|\Psi_t|}^2$,
from which you might well imagine that it follows that any variable of
interest, e.g., $X$, has the ``right'' distribution. But even if this were so
(and it is), it would be devoid of physical significance! As Einstein has
emphasized (Einstein 1953) ``Nature as a whole can only be viewed as an
individual system, existing only once, and not as a collection of
systems.''$^{11}$

While Einstein's point is almost universally accepted among physicists, it
is also very often ignored, even by the same physicists. We therefore
elaborate: What possible physical significance can be assigned to an
ensemble of universes, when we have but one universe at our disposal, the
one in which we happen to reside? We cannot perform the {\it very same\/}
experiment more than once.  But we can perform many similar experiments,
differing, however, at the very least, by location or time. In other words,
insofar as the use of probability in physics is concerned, what is relevant
is not sampling across an ensemble of universes, but sampling across space
and time within a single universe.  What is relevant is empirical
distributions---actual relative frequencies for an ensemble of actual
events.

At the expense belaboring the obvious, we stress that in order to
understand why our universe should be expected to be in \qe, it would not
be {\it sufficient\/} to establish convergence to the universal \qe\
{\it distribution\/}, even were it possible to do so. One simple consequence of
our discussion is that proofs of convergence to equilibrium for the
configuration of the universe would be of rather dubious physical
significance: What good does it do to show that an initial distribution
converges to some `equilibrium distribution' if we can attach no relevant
physical significance to the notion of a universe whose configuration is
randomly distributed according to this distribution?
In view of the implausibility of ever obtaining such a result, we are
fortunate that it is also
{\it unnecessary\/} (D\"urr, Goldstein and Zangh\`{i} 1992), as we shall now
explain.

\medskip

Two problems must be addressed, that of the meaning of the
\wf\ $\psi$ of a subsystem and that of randomness. It turns out that once
we come to grips with the first problem, the question of randomness almost
answers itself. We obtain just what we want---that $\born$ in the sense of
empirical distributions; we find (D\"urr, Goldstein and Zangh\`{i} 1992)
that in a {\it typical\/}
Bohmian universe an appearance of randomness emerges, precisely as
described by the \qf.

The term ``typical'' is used here in its mathematically precise sense: The
conclusion holds for ``almost every'' universe, i.e., with the exception of
a set of universes, or initial configurations, that is very small with
respect to a certain natural measure, namely the universal quantum
equilibrium distribution---the equivariant distribution for the universal
\BM---on the set of all universes. It is important to realize that this
guarantees that it holds for many particular universes---the overwhelming
majority with respect to the only natural measure at hand---one of which
might be ours.$^{12}$

Before proceeding to a sketch of our analysis, we would like to give a
simple example. Roughly speaking, what we wish to establish is analogous to
the assertion, following from the law of large numbers, that the {\em
relative frequency} of appearance of any particular digit in the decimal
expansion of a {\em typical} number in the interval $[0,1]$ is
$\frac{1}{10}$. In this statement two related notions appear: typicality,
referring to an a priori measure, here the Lebesgue measure, and relative
frequency, referring to structural patterns in an individual object.

	It might be objected that unlike the Lebesgue measure on $[0,1]$,
the universal quantum equilibrium measure will not in general be
uniform. Concerning this, a comment: The uniform distribution---the Lebesgue
measure on $\R^{3N}$---has no special significance for the dynamical system
defined by Bohmian mechanics. In particular, since the uniform distribution
is not equivariant, typicality defined in terms of this distribution would
depend critically on a somewhat arbitrary choice of initial time, which is
clearly unacceptable. The sense of typicality defined by the universal
\qe\ measure is independent of any choice of initial time.
\bigskip

Given a subsystem, the $x$-system, with generic configuration $x$, we may
write, for the generic configuration of the universe, $q=(x,y)$ where $y$ is
the generic configuration of the environment of the $x$-system. Similarly,
we have $Q_t=(X_t,Y_t)$ for the actual configurations at time $t$. Clearly
the simplest possibility for the wave function of the $x$-system, the
simplest function of $x$ which can sensibly be constructed from the actual
state of the universe at time $t$ (given by $Q_t$ and $\Psi_t$),  is
\be\label{cwf}
\psi_t(x)=\Psi_t(x,Y_t),
\ee
the {\it conditional \wf\/} of the $x$-system at time $t$. This is almost
all we need, almost but not quite.$^{13}$

The conditional \wf\ is not quite the right notion for the {\it effective
\wf\ \/} of a subsystem (see below; see also D\"urr, Goldstein and Zangh\`{i}
1992), since it will not in general evolve according to \se\ even when the
system is isolated from its environment. However, whenever the effective
\wf\ exists it agrees with the conditional \wf. Note, incidentally, that in
an after-measurement situation, with a system-apparatus \wf\ as in note 3,
we are confronted with the measurement problem if this \wf\ is the complete
description of the composite system after the measurement, whereas for \BM,
with the outcome of the measurement embodied in the configuration of the
environment of the measured system, say in the orientation of a pointer on
the apparatus, it is this configuration which, when inserted in
(\ref{cwf}), selects the term in the after-measurement macroscopic
superposition that we speak of as defining the ``collapsed'' system \wf\
produced by the measurement. Moreover, if we reflect upon the structure of
this superposition, we are directly led to the notion of effective \wf\
(D\"urr, Goldstein and Zangh\`{i} 1992).

Suppose that
\be\label{ewf}
\Psi_t(x,y)=\psi_t(x)\Phi_t(y)+\Psi_t^\perp(x,y),
\ee
where $\Phi_t$ and $\Psi_t^\perp$ have macroscopically disjoint
$y$-supports. If
\be\label{supp}
Y_t\in \mbox{supp}\,{\Phi_t}
\ee
we say that $\psi_t$ is the {\it \ewf\/} of the $x$-system at time $t$.
Note that it follows from (\ref{supp}) that
$\Psi_t(x,Y_t)=\psi_t(x)\Phi_t(Y_t)$, so that the effective
\wf\ is unambiguous, and indeed agrees with the conditional \wf\ up to an
irrelevant constant factor.

We remark that it is the relative stability of the macroscopic disjointness
employed in the definition of the
\ewf, arising from what are nowadays often called mechanisms of
decoherence---the destruction of the coherent spreading of the wave
function due to dissipation, the effectively irreversible flow of
``phase information'' into the (macroscopic) environment---which accounts
for the fact that the \ewf\ of a system obeys
\Sc's equation for the system alone whenever this system is isolated. One
of the best descriptions of the mechanisms of decoherence, though not the
word itself, can be found in  Bohm's 1952 ``hidden variables''
paper (Bohm 1952). We wish to emphasize, however, that while decoherence
plays a crucial role in the very formulation of the various interpretations
of \qt\ loosely called decoherence theories, its role in Bohmian mechanics
is of a quite different character: For Bohmian mechanics decoherence is
purely phenomenological---it plays no role whatsoever in the formulation
(or interpretation) of the theory itself.$^{14}$

An immediate consequence (D\"urr, Goldstein and Zangh\`{i} 1992) of
(\ref{cwf}) is the {\it fundamental conditional probability formula\/}:
\be\label{fpf} {\P}\bigl(X_t \in dx \bigm| Y_t\bigr)=|\psi_t(x)|^2\,dx, \ee
where $\P(dQ)={|\Psi_0(Q)|}^2\,dQ$.

Now suppose  that at time $t$ the $x$-system consists
itself of many identical subsystems
$x_1,\dots,x_M$, each one having \ewf\ $\psi$ (with respect to  coordinates
relative to suitable frames).
Then (D\"urr, Goldstein and Zangh\`{i} 1992) the \ewf\ of the $x$-system is
the product \wf\
\be\label{pf}
\psi_t(x)=\psi(x_1)\cdots\psi(x_M).
\ee

Note that it follows from (\ref{fpf}) and (\ref{pf}) that the
configurations of these subsystems are independent identically distributed
random variables with respect to the initial universal quantum equilibrium
distribution $\P$ conditioned on the environment of these subsystems. Thus
the law of large numbers can be applied to conclude that the empirical
distribution of the configurations $X_1,\dots,X_M$ of the subsystems will
typically be $|\psi(x)|^2$---as demanded by the quantum formalism. For
example, if $|\psi|^2$ assigns equal probability to the events ``left'' and
``right,'' typically about half of our subsystems will have configurations
belonging to ``left'' and half to ``right.'' Moreover (D\"urr, Goldstein
and Zangh\`{i} 1992), this conclusion applies as well to a collection of
systems at possibly different times as to the equal-time situation
described here.$^{15}$

It also follows (D\"urr, Goldstein and Zangh\`{i} 1992) from the formula
(\ref{fpf}) that a typical universe embodies {\it absolute uncertainty\/}:
the impossibility of obtaining more information about the present
configuration of a system than what is expressed by the \qe\ hypothesis.
In this way, ironically, \BM\ may be regarded as providing a sharp
foundation for and elucidation of Heisenberg's uncertainty principle.

\section{What is a Bohmian Theory?}

\BM, the theory defined by eqs.(\ref{eqsc})  and (\ref{velo}), is not Lorentz
invariant, since (\ref{eqsc}) is a nonrelativistic equation, and, more
importantly, since the right hand side of (\ref{velo}) involves the
positions of the particles at a common (absolute) time. It is also
frequently asserted that \BM\ cannot be made Lorentz invariant, by which it
is presumably meant that no Bohmian theory---no theory that could be
regarded somehow as a natural extension of \BM---can be found that is
Lorentz invariant.  The main reason for this belief is the manifest
nonlocality of \BM\ (Bell 1987).  It must be stressed, however, that
nonlocality has turned out to be a fact of nature: nonlocality must be a
feature of any physical theory accounting for the observed violations of
Bell's inequality. (See Bell 1987 and the contributions of Maudlin and
Squires to this volume.)

A serious difficulty with the assertion that \BM\ cannot be made Lorentz
invariant is that what it actually means is not at all clear, since this
depends upon what is to be understood by a Bohmian theory. Concerning this
there is surely no uniformity of opinion, but what we mean by a {\it
Bohmian theory\/} is the following:

\smallskip

\noindent 1) A Bohmian theory should be based upon a clear ontology, the
primitive ontology, corresponding roughly to Bell's local beables. This
primitive ontology is what the theory is fundamentally about. For the
nonrelativistic theory that we have been discussing, the primitive ontology
is given by particles described by their positions, but we see no
compelling reason to insist upon this ontology for a  relativistic
extension of \BM.

Indeed, the most obvious ontology for a bosonic field theory is a field
ontology, suggested by the fact that in standard quantum theory, it is the
field configurations of a bosonic field theory that plays the role
analogous to that of the particle configurations in the particle theory.
However, we should not insist upon the field ontology either. Indeed, Bell
(Bell 1987, 173--180) has proposed a Bohmian model for a quantum field
theory involving both bosonic and fermionic quantum fields in which the
primitive ontology is associated only with fermions---with no local
beables, neither fields nor particles, associated with the bosonic quantum
fields. Squires (contribution in this volume) has made a similar proposal.

While we insist that a Bohmian theory be based upon some clear ontology, we
have no idea what the appropriate ontology for relativistic physics
actually is.

\smallskip

\noindent 2) There should be a quantum state, a \wf, that evolves according
to the unitary quantum evolution and whose role is to somehow generate the
motion for the variables describing the primitive ontology.

\smallskip

\noindent 3) The predictions should agree (at least
approximately) with those of orthodox quantum theory---at least to the
extent that the latter are unambiguous.

\medskip

This description of what a Bohmian theory should involve is admittedly
vague, but greater precision would be inappropriate. But note that, vague
as it is, this characterization clearly separates a Bohmian theory from an
orthodox quantum theory as well as from the other leading alternatives to
Copenhagen orthodoxy: The first condition is not satisfied by the
decoherent or consistent histories (Griffiths 1984, Omn\`es 1988, Gell-Mann
and Hartle 1993) formulations while with the spontaneous localization
theories the second condition is deliberately abandoned (Ghirardi, Rimini
and Weber 1986 and Ghirardi, Pearle and Rimini 1990). With regard to the
third condition, we are aware that it is not at all clear what should be
meant by even an orthodox theory of quantum cosmology or gravity, let alone
a Bohmian one.  Nonetheless, this condition places strong constraints on
the form of the guiding equation.

Furthermore, we do not wish to suggest here that the ultimate theory is
likely to be a Bohmian theory, though we do think it very likely that if
the ultimate theory is a quantum theory it will in fact be a Bohmian
theory.

Understood in this way, a Bohmian theory is merely a quantum theory with a
coherent ontology. If we believe that ours is a quantum world, does this
seem like too much to demand? We see no reason why there can be no Lorentz
invariant Bohmian theory. But if this should turn out to be impossible, it
seems to us that we would be wiser to abandon Lorentz invariance before
abandoning our demand for a coherent ontology.

\section*{Acknowledgments} We are grateful to Karin Berndl, James Cushing,
and Eugene Speer for valuable suggestions. This work was
supported in part by NSF Grant No. DMS--9504556, by the DFG, and by the INFN.

\begin{center} {\bf Notes\/}
\end{center}

\noindent$^{1}$\indent When a magnetic field is present, the gradients in the
equations must be understood as the covariant derivatives involving the
vector potential.  If $\psi$ is spinor-valued, the bilinear forms appearing
in the numerator and denominator of (\ref{velo}) should be understood as
spinor-inner-products. (See the discussion of spin in Section 3.)  For
indistinguishable particles, it follows from a careful analysis (D\"urr,
Goldstein and Zangh\`{i} 1996) of the natural configuration space, which will
no longer be $\R^{3N}$, that when the \wf\ is represented in the usual way,
as a function on $\R^{3N}$, it must be either symmetric or antisymmetric
under permutations of the labeled position variables.  Note in this regard
that according to orthodox quantum mechanics, the very notion of
indistinguishable particles seems to be grounded on the nonexistence of
particle trajectories. It is thus worth emphasizing that with \BM\ the
classification of particles as bosons or fermions emerges naturally from
the very existence of trajectories.

\noindent$^{2}$\indent  The contortions required to deal
with spin in the spirit of the quantum potential are particularly striking
(Bohm and Hiley 1993, Holland 1993).

\noindent$^{3}$\indent This is really no assumption at all, since the outcome
should ultimately be converted to digital form, whatever its initial
representation may be.

\noindent$^{4}$\indent In the simplest such situation the unitary evolution for
the \wf\ of the composite system carries the initial \wf\
$\Psi_i=\psi\ot\Phi_0$ to the final \wf\ $\Psi_f=\sum_\a\psia\ot\Phi_\a$,
where $\Phi_0$ is the ready apparatus \wf, and $\Phi_\a$ is the apparatus
\wf\ corresponding to outcome $\a$. Then integrating ${|\Psi_f|}^2$ over
supp$\,\Phi_\a$, we immediately arrive at (\ref{prob}).

\noindent$^{5}$\indent Operators as observables also naturally convey
information about the system's \wf\ after the experiment. For example, for
an ideal measurement, when the outcome
is $\a$ the \wf\ of the system after the experiment is (proportional to)
$P_{\H_\a}\psi$. We shall touch briefly upon this collapse of the \wf,
i.e., the projection postulate, in Section 4, in connection with the notion
of the effective \wf\ of a system.

\noindent$^{6}$\indent Even speaking of the observable $A$ as having value
$\lambda_\a$ when the system's \wf\ is in $\H_\a$, i.e., when this \wf\ is
an eigenstate of $A$ of eigenvalue $\lambda_\a$, insofar as it suggests
that something peculiarly quantum is going on when the \wf\ is not an
eigenstate whereas in fact there is nothing the least bit peculiar about
the situation, perhaps does more harm than good.

\noindent$^{7}$\indent It also applies to the spontaneous collapse models
(Ghirardi, Rimini and Weber 1986 and Ghirardi, Pearle and Rimini 1990), the
interpretation of which (see, e.g., Albert 1992, 92--111) is often marred
by \nrao. See, however, Bell's presentation of GRW (Bell 1987, 205) for an
illuminating exception, as well as Ghirardi, Grassi and Benatti
1995 and the contribution of Ghirardi and  Grassi to this volume.

\noindent$^{8}$\indent Here we use the usual notation $\left(\begin{array}{c}
a\\b\end{array}\right)\otimes \phi_0$ for $\left(\begin{array}{c}
a\phi_0\\b\phi_0\end{array}\right)$.

\noindent$^{9}$\indent We should probably distinguish two senses of
``primitive'': i) the {\it strongly primitive\/} variables, which describe
what the theory is fundamentally {\it about\/}, and ii) the {\it weakly
primitive\/} variables, the basic variables of the theory, those which
define the complete state description.  The latter may either in fact be
strongly primitive, or, like the electromagnetic field in classical
electrodynamics, they may be required in order to express the laws which
govern the behavior of the strongly primitive variables in a simple and
natural way. While this probably does not go far enough---we should further
distinguish those weakly primitive variables which, like the velocity, are
functions of the trajectory of the strongly primitive variables, and those,
again like the electromagnetic field, which are not---these details are not
relevant to our present purposes, so we shall ignore these distinctions.

\noindent$^{10}$\indent As an illustration of the pitfalls in trying to
establish convergence to quantum equilibrium, a recent attempt of Valentini
(Valentini 1991) is instructive. Valentini's argument is based on a
``subquantum $H$-theorem,'' $d\bar H /dt\leq0$, that is too weak to be of
any relevance (since, for example, the inequality is not strict). The
$H$-theorem is itself not correctly proven---it could not be since it is in
general false. Even were the $H$-theorem true, correctly proven, and
potentially relevant, the argument given would still be circular, since in
proceeding from the $H$-theorem to the desired conclusion, Valentini finds
it necessary to invoke ``assumptions similar to those of classical
statistical mechanics,'' namely that (Valentini 1992, 36) ``the system is
`sufficiently chaotic','' which more or less amounts to assuming the very
mixing which was to be derived.

\noindent$^{11}$\indent For a rather explicit example of the failure to
appreciate this point, see Albert 1992, 144: ``And the statistical
postulate \dots\ can be construed as stipulating something about the {\it
initial conditions\/} of the universe; it can be construed (in the
fairy-tale language, say) as stipulating that what God did when the
universe was created was first to choose a wave function for it and
sprinkle all of the particles into space in accordance with the
quantum-mechanical probabilities, and then to leave everything alone,
forever after, to evolve deterministically. And note that just as the
one-particle postulate can be derived \dots\ from the two-particle
postulate, {\it all\/} of the more specialized statistical postulates will
turn out to be similarly derivable from {\it this\/} one.''
(Note that an initial sprinkling in accordance with the quantum-mechanical
probabilities need remain so only if ``quantum-mechanical probabilities''
is understood as referring to the quantum equilibrium distribution for the
configuration of the entire universe rather than to empirical distributions
for subsystems arising from this configuration. Note also that the analogy
with the relationship between the one-particle and the two-particle
postulates also requires that the universal ``statistical postulate'' be
understood in this way.)

\noindent$^{12}$\indent It is important to realize that an appeal to typicality
is unavoidable if we are to explain why the universe is at present in
quantum equilibrium. This is because our analysis also demonstrates that
there is a set $B$ of initial configurations, a set of nonvanishing
Lebesgue measure, that evolve to present configurations violating the \qe\
hypothesis and hence the quantum formalism. This set cannot be wished away
by any sort of mixing argument. Indeed, if, as is expected, mixing holds on
the universal level, then this set $B$ should be so convoluted as to be
indescribable without a specific reference to the universal dynamics and
hence cannot be dismissed as unphysical without circularity.

\noindent$^{13}$\indent For particles with spin, (\ref{cwf}) should be replaced
by $\Psi_t(x,Y_t)=\psi_t(x)\ot\Phi_t(Y_t)$.  In particular, for particles
with spin, not every subsystem has a conditional \wf.

\noindent$^{14}$\indent However, decoherence is important for a serious
discussion of the emergence of Newtonian mechanics as the description of
the macroscopic regime for Bohmian mechanics, leading to the picture of a
macroscopic Bohmian particle, in the classical regime, guided by a
macroscopically well-localized wave packet with a macroscopically sharp
momentum moving along a classical trajectory. It may, indeed, seem somewhat
paradoxical that the gross features of our world should appear classical
because of interaction with the environment and the resulting wave function
entanglement (Joos and Zeh 1985, Gell-Mann and Hartle 1993), the
characteristic quantum innovation (Schr\"odinger 1935).

\noindent$^{15}$\indent It should not be necessary to say that we do not claim
to have established the impossibility---but rather the atypicality---of
quantum nonequilibrium. On the contrary, as we have suggested in the first
reference of D\"urr, Goldstein and Zangh\`{i} 1992, 904, ``the reader may
wish to explore quantum nonequilibrium. What sort of behavior would emerge
in a universe which is in quantum nonequilibrium?''  Concerning this, we
wish to note that despite what is suggested by the misuse of ensembles for
the universe as a whole and the identification of the physical universal
convergence of configurations to those characteristic of \qe\ with the
expected convergence of universal measures, of $P\to|\Psi|^2$, \qe\ is not
an attractor, and no ``force'' pushes the universal configuration to one of
\qe. Rather, any transition from quantum nonequilibrium to \qe\ would be
entropic and time-symmetric---driven indeed primarily by measure-theoretic
effects, by the fact that the set of \qe\ configurations is vastly larger
than the set of configurations corresponding to quantum
nonequilibrium---just as is the convergence to thermodynamic
equilibrium. (For some speculations on the possible value of quantum
nonequilibrium, see the contribution of Valentini to this volume.)

\newpage

\begin{center} {\bf References\/}

[For D\"urr, Goldstein, and Zanghi]
\end{center}

\parindent=-25pt

Albert, D.Z. (1992), {\it Quantum Mechanics and Experience\/}. Cambridge,
MA, Harvard University Press.

Bell, J.S. (1987), {\it Speakable and Unspeakable in
Quantum Mechanics\/}. Cambridge, Cambridge University Press.

Bohm, D. (1952), ``A Suggested Interpretation of the Quantum Theory in Terms
of `Hidden' Variables, I and II,'' {\it Physical Review} {\bf 85}, 166--193.

Bohm, D. (1953), ``Proof that Probability Density Approaches $|\psi|^2$ in
Causal Interpretation of Quantum Theory,'' {\it Physical Review\/} {\bf
89}, 458--466.

Bohm, D. (1980), {\it Wholeness and the Implicate Order\/}. New
York, Routledge.

Bohm, D. and Hiley, B.J. (1993), {\it The Undivided Universe: An
Ontological Interpretation of Quantum Theory\/}. London, Routledge \& Kegan
Paul.

Daumer, M., D\"urr, D., Goldstein, S. and Zangh\`{i}, N. (1994), ``Scattering
and the Role of Operators in Bohmian Mechanics'' in M.~Fannes, C.~Maes, and
A.~Verbeure (eds.), {\it On Three Levels}. New York, Plenum Press, pp. 331-338.

Daumer, M., D\"urr, D., Goldstein, S. and Zangh\`{i}, N. (1996), ``On the
Role of Operators in Quantum Theory,'' in preparation.

Davies, E.B. (1976), {\it Quantum Theory of Open Systems\/}. London,
Academic Press.

D\"urr, D., Goldstein, S. and Zangh\`{i}, N. (1992), ``Quantum Equilibrium and
the Origin of Absolute Uncertainty,'' {\it Journal of Statistical
Physics} {\bf 67}, 843--907; ``Quantum Mechanics, Randomness, and
Deterministic Reality,''{\it Physics Letters A} {\bf 172}, 6--12.

D\"urr, D., Goldstein, S. and Zangh\`{i}, N. (1996), ``Bohmian Mechanics,
Identical Particles, Parastatistics and Anyons,'' in preparation.

Einstein, A. (1953), in {\it Scientific Papers Presented to Max
Born\/}. Edinburgh, Oliver \& Boyd, pp. 33--40.

Feynman, R.P., Leighton, R.B. and Sands, M. (1963), {\it The Feynman
Lectures on Physics, I\/}. New York, Addison-Wesley.

Gell-Mann, M. and Hartle, J.B. (1993), ``Classical Equations for Quantum
Systems,'' {\it Physical Review D\/} {\bf 47}, 3345--3382.

Ghirardi, G.C., Rimini, A. and Weber, T. (1986), ``Unified Dynamics for
Microscopic and Macroscopic Systems,'' {\it Physical Review D\/} {\bf 34},
470--491.

Ghirardi, G.C., Pearle, P. and Rimini, A. (1990), ``Markov Processes in
Hilbert Space and Continuous Spontaneous Localization of Systems of
Identical Particles,'' {\it Physical Review A\/} {\bf 42}, 78--89.

Ghirardi, G.C., Grassi, R. and Benatti, F. (1995), ``Describing the
Macroscopic World: Closing the Circle within the Dynamical Reduction
Program,'' {\it Foundations of Physics\/} {\bf 23}, 341--364.

Griffiths, R.B. (1984), ``Consistent Histories and the Interpretation of
Quantum Mechanics,'' {\it Journal of Statistical Physics\/} {\bf 36\/},
219--272.

Holland, P.R. (1993), {\it The Quantum Theory of Motion\/}. Cambridge,
Cambridge University Press.

Mott, N.F. (1929), {\it Proceedings of the Royal Society A\/} {\bf 124\/}, 440.

Omn\`es, R. (1988), ``Logical Reformulation of Quantum Mechanics I,''
{\it Journal of Statistical Physics\/} {\bf 53\/}, 893--932.

Schr\"odinger, E. (1935), ``Die gegenw\"artige Situation in der
Quantenmechanik,'' {\it Die Naturwissenschaften\/} {\bf 23\/}, 807--812,
824--828, 844-849. [Also appears in translation as ``The Present Situation
in Quantum Mechanics,'' in Wheeler and Zurek 1983, 152--167.]

Valentini, A. (1991), ``Signal-Locality, Uncertainty, and
the Subquantum $H$-Theorem. I,'' {\it Physics Letters A\/} {\bf 156}, 5--11.

Valentini A. (1992), {\it On the Pilot-Wave Theory of Classical, Quantum
and Subquantum Physics\/}. Ph.D. thesis, International School for Advanced
Studies, Trieste.

Wheeler, J.A. and Zurek, W.H. (eds.)(1983), {\it Quantum Theory and
Measurement\/}. Princeton,  Princeton University Press.

\end{document}